\documentclass[a4paper,11pt]{article}
\usepackage{ieee6x9}
\usepackage{times,epsfig} 
\title{Structure induction by lossless graph compression}
\author{Leonid Peshkin\\
\it Center for Biomedical Informatics \\
\it Harvard Medical School \\
\it Boston, MA 02115, USA\\
\it pesha@hms.harvard.edu}
\date{}
\begin{document}
\maketitle
\thispagestyle{empty}
\begin{abstract}

This work is motivated by the necessity to automate the discovery of structure in vast and ever-growing collection of relational data commonly represented as graphs, for example genomic networks. 
A novel algorithm, dubbed {\em Graphitour}, for structure induction by lossless graph compression is presented and illustrated by a clear and broadly known case of nested structure in a DNA molecule. 
This work extends to graphs some well established approaches to grammatical inference previously applied only to strings. The bottom-up graph compression problem is related to the maximum cardinality (non-bipartite)  maximum cardinality matching problem.
The algorithm accepts a variety of graph types including directed graphs and graphs with labeled nodes and arcs. 
The resulting structure could be used for representation and classification of  graphs. 
\end{abstract}

\section{Introduction}

The explosive growth of relational data, for example data about genes, drug molecules and proteins, their functions and interactions, necessitates efficient mathematical algorithms and software tools to extract meaningful generalizations. There is a large body of literature on the subject coming from a variety of disciplines from Theoretical Computer Science to Computational Chemistry. 
However, one fundametal issue has so far remained unaddressed. Given a multi-level nested network of relations, such as a complex molecule, or a protein-protein interaction network, how can its structure be inferred from first principles. This paper is meant to fill this surprising gap in automated data processing. 

Let us illustrate the purpose of this method through the description of DNA molecular structure, the way most of us learned it from a textbook or in class. 
\begin{itemize}
\item The DNA molecule is a double chain made of four kinds of nucleotides: A, T, G, C; 
\item Each of these is composed of two parts: one part---backbone---is identical among all the nucleotides (neglecting the difference between ribose and 2'-deoxyribose), another---heterocyclic base---is nucleotide-specific;
\item The backbone consists of sugar and phosphate;
\item The heterocyclic bases (C,T-pyrimidines; A,G-purines) all contain a pyrimidine ring; 
\item The components can be further reduced to individual atoms and covalent bonds. 
\end{itemize}
This way of description is not unique, and may be altered according to the desired level of detail, but crucially, it is a hieararchical description of the whole as a structure built from identifiable and repetitive subcomponents. The picture of this beautiful multi-level hierarchy has emerged after years of bio-chemical discovery by scientists who gradually applied their natural abstraction and generalization abilities. Hence, structural elements in this hierarchy also make functional sense from bio-chemical point of view.


\begin{figure}[tb]
\centerline{
\epsfig{file=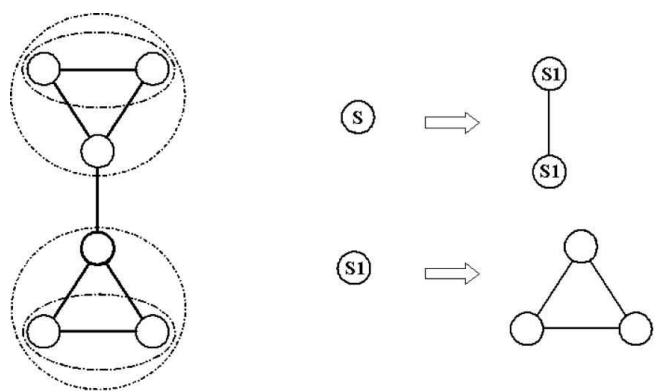,height=5cm}}
\caption{A graph with repetative structure and a corresponding grammar.}\label{fig40}
\end{figure}

The properties of hierarchical description are formally well-studied and applied in other scientific domains, such as linguistics and computer science. It is viewed as the result of a rule-driven generative process, which combines a finite set of undecomposable elements---{\it terminal symbols}---into novel complex objects---{\it non-terminal symbols}, which can be combined in turn to produce the next level of description. The rules and symbols on which the process operates are determined by a {\it grammar} and the process itself is termed a {\it grammatical derivation}. In the case of the DNA molecule above, the chemical elements correspond to terminal symbols. They are assembled into non-terminal symbols, i.e. compounds, according to some set of production rules defined by chemical properties. 

Now, imagine receiving an alternative description of the same object, stripped off of any domain knowledge and context, simply as an enormous list of objects and binary relations on objects, corresponding to thousands of atoms and covalent bonds. Such a list would remain completely incomprehensible to a human mind, along with any repetitive hierarchical structure present in it. Discovering a hierarchy of nested elements without any prior knowledge of a kind, size and frequency of these constitutes a formidable challenge. 

Remarkably, this is precisely the challenge which is undertaken by contemporary scientists trying to make sense of data, mounting up from small fragments, like protein interaction networks, regulatory and metabolic pathways, small molecule repositories, homology networks, etc. Our goal is to be able to approach such  tasks in an automated fashion. 

Figure~\ref{fig40} illustrates the kind of induction we describe in this paper on a trivial example. We will use this as a running example throughout the paper, leaving more rigorous mathematical formulation out for the purpose of clarity and wider accessibility. 
To the left is a graph which contains repetitive structure. Let us imagine for a moment that the human researcher is not smart enough to comprehend a 6-node graph and find an explanatory layout. Thus, we would want to automatically translate such a graph into the graph grammar on the right. The graph grammar consist of two productions. The first expands a starting representation---a degenerate graph of a single node "S"---into a graph connecting two nodes of the same type "S1". The second additionally defines a node "S1" as a fully connected triple. 

\begin{table}[thb]
\caption{Generalized string compression algorithm.}
\rule{6in}{1pt}
\begin{tabbing}
111\=111\=1111\=11\=111\= \kill
  \> {\bf Input:} initial string;\\
  \> {\bf Initialize:} empty grammar;\\
  \> {\bf Loop:}  \\
  \> \> Make a single left-to-right tour of the string, collecting sub-string statistics; \\
  \> \> Introduce a new non-terminal symbol (compound) into the grammar,\\ 
   \> \> \>  which  agglomerates some sub-string of terminal and non-terminal symbols;\\
  \> \> Substitute all occurrences of a new compound; \\
  \> {\bf Until} no compression possible;\\
  \> {\bf Output:} compressed string \& induced grammar; 
\end{tabbing}
\rule{6in}{1pt}
\label{alg_str}
\end{table}

Formal description of the relational data of such kind is known as graphs, while the hierarchical nested structures of such kind are described by graph grammars. It is outside the scope of this paper to survey a vast literature in the field of graph grammars; please refer to a book by G. Rozenberg~\cite{Rozenberg97} for extensive overview. It suffices to say that this field is mostly concerned with the transformation of the graphs, or parsing, i.e. explaining away a graph according to some known graph grammar, rather than with inducing such grammar from raw data. 

The closest work related to the ideas presented here is due to D. Cook, L. Holder and their colleagues (e.g. see~\cite{Cook94} and several follow-up papers). Their work however is not concerned with inducing a structure from given graph data. Rather, they induce a flat, context-free grammar, possibly with recursion, which is not only capable of, but is also bound to, generate objects not included in the original data. Thus, their approach defies the relation to compression exploited here. Moreover, the authors present the negative result of running their {\sc subdue} algorithm on just the kind of biological data we successfully use in this paper. Another remotely similar work is by Stolke~\cite{Stolke93} in application to inducing hidden Markov models. 
There are many other works attempting to induce structure from relational data or compress graphs, but none seem to relate closely to the method considered here. 

Our method builds on the parallels between understanding and compression. Indeed, to understand some phenomenon from the raw data means to find some repetitive pattern and hierarchical structure, which in turn could be exploited to re-encode the data in a compact way. This work extends to graphs some well established approaches to grammatical inference previously applied only to strings. 
Two methods particularly worth mentioning in this context for grammar induction on sequences are {\it Sequitour}~\cite{Sequitour} and {\sc adios}~\cite{ADIOS}. We also take inspiration from a wealth 
of sequence compression algorithms, often unknowingly run daily by all computer users in a form of archival software like {\em pkzip} in Unix  or {\em expand} for Mac OS X.

Let us briefly convey the intuition behind such algorithms, many of which are surveyed by Lehman and Shelat~\cite{Lehman02}. 
Although quite different in detail, all algorithms share common principles and have very similar compression ratio and computational complexity bounds. 
First, one has to remember that all such compression/discovery algorithms are bound to be heuristics, since finding the best compression is related to the so-called Kolmogorov complexity and is provably hard~\cite{Lehman02}. These heuristics are in turn related to the MDL (Minimum Description Length) principle, and work in the way described by Table~\ref{alg_str}. 
\begin{table}[hbt]
\caption{{\it Graphitour}: graph compression algorithm.}
\rule{6in}{1pt}
\begin{tabbing}
111\=111\=1111\=11\=111\= \kill
  \> {\bf Input:} initial graph: lists of (labeled) nodes and edges;\\
  \> {\bf Initialize:} empty graph grammar, empty edge lexicon;\\
  \> {\bf Build edge lexicon:} make a single tour of the graph,  register the type of each edge\\
  \> \> according to the edge label and types of its end nodes; collect type statistics; \\
  \> {\bf Loop:}  \\
 \> \> {\bf Loop through edge types} \\
  \> \> \> For a sub-graph induced by each edge type solve an instance of {\em maximum}\\
  \> \> \>  {\em cardinality matching} problem, which yields a number and list of edges\\
  \> \> \>  that could be abstracted; \\ 
  \> \> {\bf Pick an edge type} which corresponds to the highest count;\\
  \> \> {\bf Introduce a new hyper-node} for the chosen edge type into the graph grammar; \\ 
  \> \> {\bf Loop} through edge occurrences of a chosen type throughout the graph;\\
  \> \> \> Substitute an occurrence of the edge type with a hyper-node; \\ 
  \> \> \>  {\bf Loop} through all edges incident to the end nodes of a given edge; \\
  \> \> \> \> Substitute edges and introduce new edge types into edge lexicon;\\
  \> {\bf Until} no compression possible;\\
  \> {\bf Output:} compressed graph \& induced graph grammar; 
\end{tabbing}
\rule{6in}{1pt}
\label{alg_grph}
\end{table} 
Naturally, the difference is in how exactly statistics are used to pick which substring will be substituted by a new compound symbol. In some cases, a greedy strategy is used (see e.g. Apostolico \& Lonardi~\cite{Apostolico}), i.e. the substitution which will maximally reduce the size of the encoding at the current step is picked; in other cases, a simple first-come-first-served principle is used and any repetition is immediately eliminated (see e.g. Nevill-Manning \& Witten~\cite{Sequitour}). 

Extending these methods to a graph structure turns out to be non-trivial for several reasons. First, maintaining a lexicon of strings and looking up entries is quite different for graphs. Second, directly extending the greedy approach~\cite{Apostolico} fails due to inherent non-linear entity interactions in a graph.

\section{Algorithm}

The algorithm we present is dubbed "{\em Graphitour}" to acknowledge a strong influence of the Sequitour algorithm~\cite{Sequitour} for discovering the hierarchical structure in sequences. In particular, {\em Graphitour} just like Sequitour restricts the set of candidate structures to bigrams, which allows it to eliminate a rather costly lexicon lookup procedure, as well as sub-graph isomorphism resolution, which is NP-hard.  
	Note that {\em Graphitour} in principle admits labeled nodes and edges, particularly since unlabeled graphs turn into labeled at the first iteration of the algorithm. 

\begin{figure}[t]
\centerline{\includegraphics[height=5.5cm]{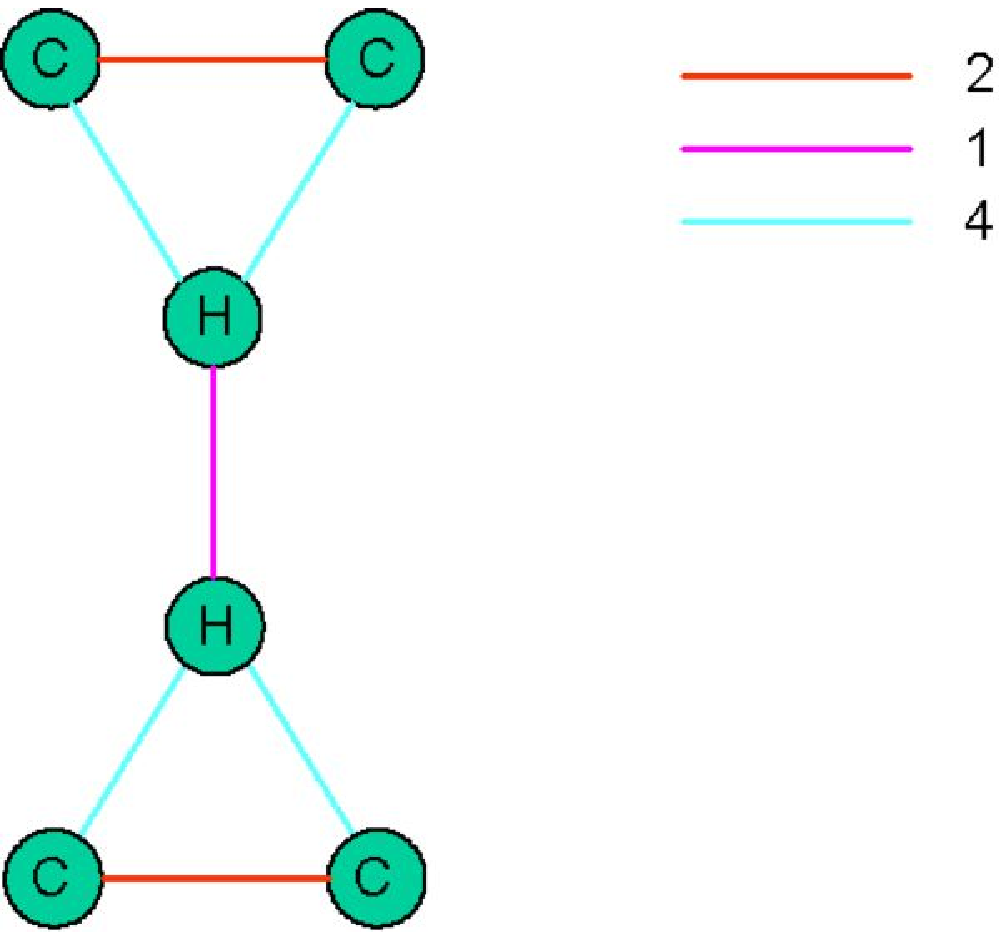}\hspace{2 cm}\includegraphics[height=5.5cm]{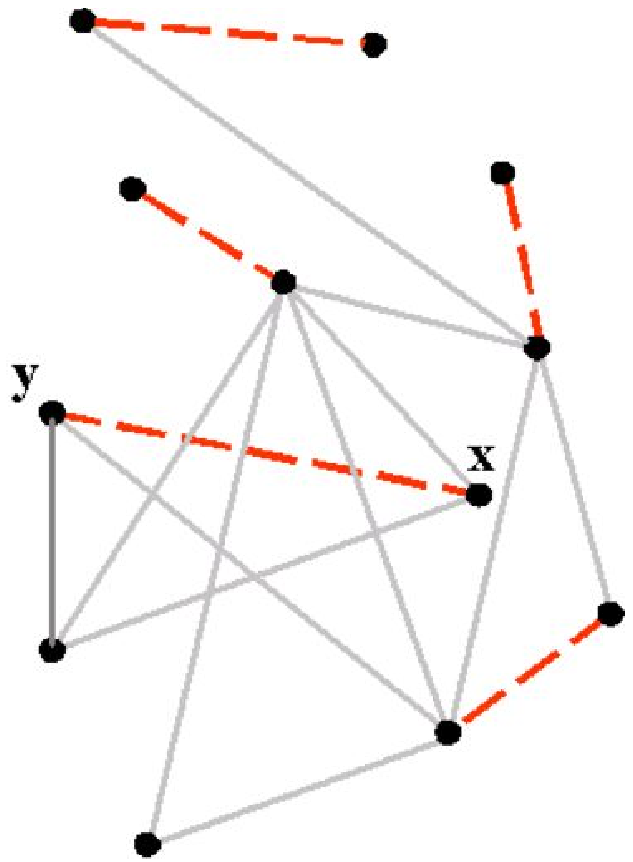}}
\caption{{\bf left:} A simple graph with labeled nodes corresponding to an imaginary small molecule and a corresponding sample edge lexicon with type frequency counts.
{\bf right:} Non-bipartite maximum cardinality matching.}\label{fig49}
\end{figure}

The algorithm is presented in Table~\ref{alg_grph} and best understood through the illustration of every step on a trivial example of Figure~\ref{fig49} as described below.
\begin{figure}[b]
\centerline{\includegraphics[height=5.3cm]{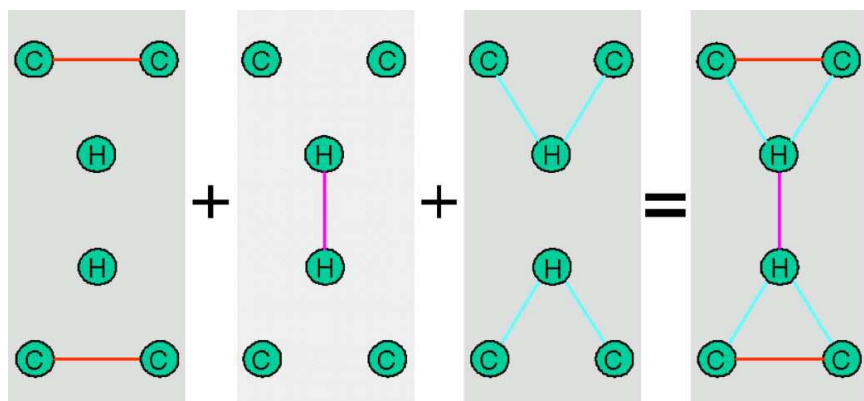}}
\caption{A graph decomposition into subgraphs induced by edges of different type.}\label{fig54}
\end{figure}
Figure~\ref{fig49} presents an imaginary small molecule with two types of nodes-atoms: "H" and "C". Each edge is typed according to its end nodes. The first iteration is to tour the graph making a lexicon of edge types and corresponding frequencies. 
Figure~\ref{fig49} shows three kinds of edges we observe: "C-C", "C-H" and "H-H" with corresponding counts. 

\begin{figure}[th]
\centerline{\includegraphics[height=5.3cm]{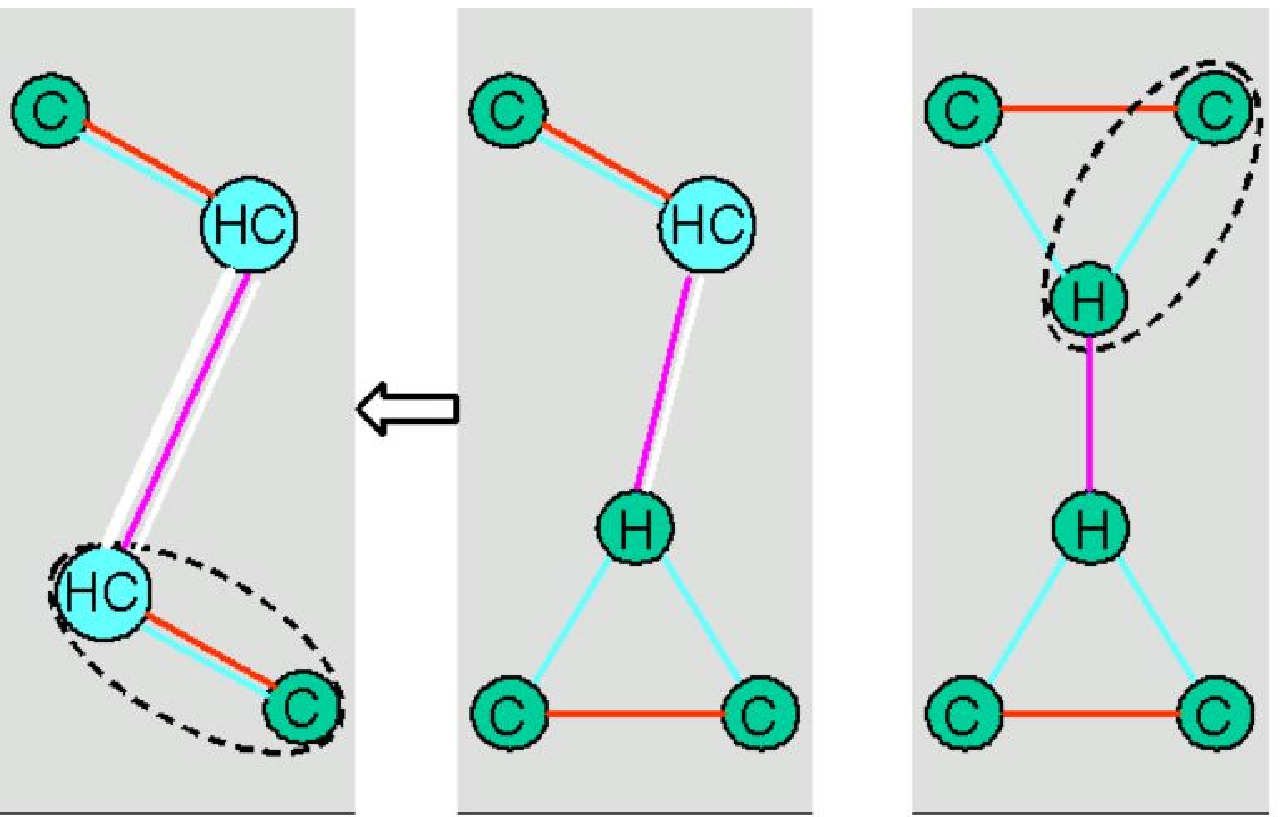}}
\caption{A sequence of abstraction steps, right to left: a new entry is made into the lexicon, called "HC" which corresponds to an abstracted edge.}\label{fig60}
\end{figure}

The next step is to select which kind of edge will be abstracted into a (hyper)node corresponding to a new compound element. A greedy choice would have picked the most frequent type of edge, disregarding potential conflicts due to shared node interactions. Instead {\em Graphitour} proceeds by decomposing the graph into three graphs each containing only edges of one type, which is illustrated by Figure~\ref{fig54},
and counting how many edges of each type could be abstracted, taking into account node interactions. 

\newpage
One of the key contributions of the algorithm described here is in relating the graph grammar induction to the {\em maximum cardinality non-bipartite matching problem} for which there is a polynomial complexity algorithm due to Gabow~\cite{Gabow73}. 
The problem is formulated as follows: {\em given a graph, find the largest set of edges, such that no two edges share a node.} Figure~\ref{fig49}.right illustrates this on an arbitrary non-bipartite graph. Dashed edges represent a valid matching. The description and implementation of the algorithm is quite technically involved and does not belong to the scope of this paper.  Note that while applying Gabow's elegant algorithm essentially allows us to extend the horizon from one-step-look-ahead greedy to two-step-look-ahead, the bottom-up abstraction process still remains a greedy heuristic. 
Only one partial graph of Figure~\ref{fig54}, containg "C-H" edges, makes a non-degenerate case for a maximal non-bipartite matching since edges interact at the node "H", obviously returning only two out of four edges subject to abstraction, marked with an oval shape on the figure. A new entry is made into the lexicon, called "HC" which corresponds to an abstracted edge, as illustrated by two steps going right to left in Figure~\ref{fig60}. 

The leftmost sub-figure of Figure~\ref{fig60} is an input for the next iteration of the algorithm, which detects the new repetitive compound: edge "HC-C", which is abstracted again, ultimately producing a graph grammar shown in the Figure~\ref{fig40} on the right. Note that as the algorithm iterates through the graph, intermediate structures get merged with the higher level structures. 

\section{Results}

\begin{figure}[bht]
\centerline{\includegraphics[width=12.5cm,height=9.cm]{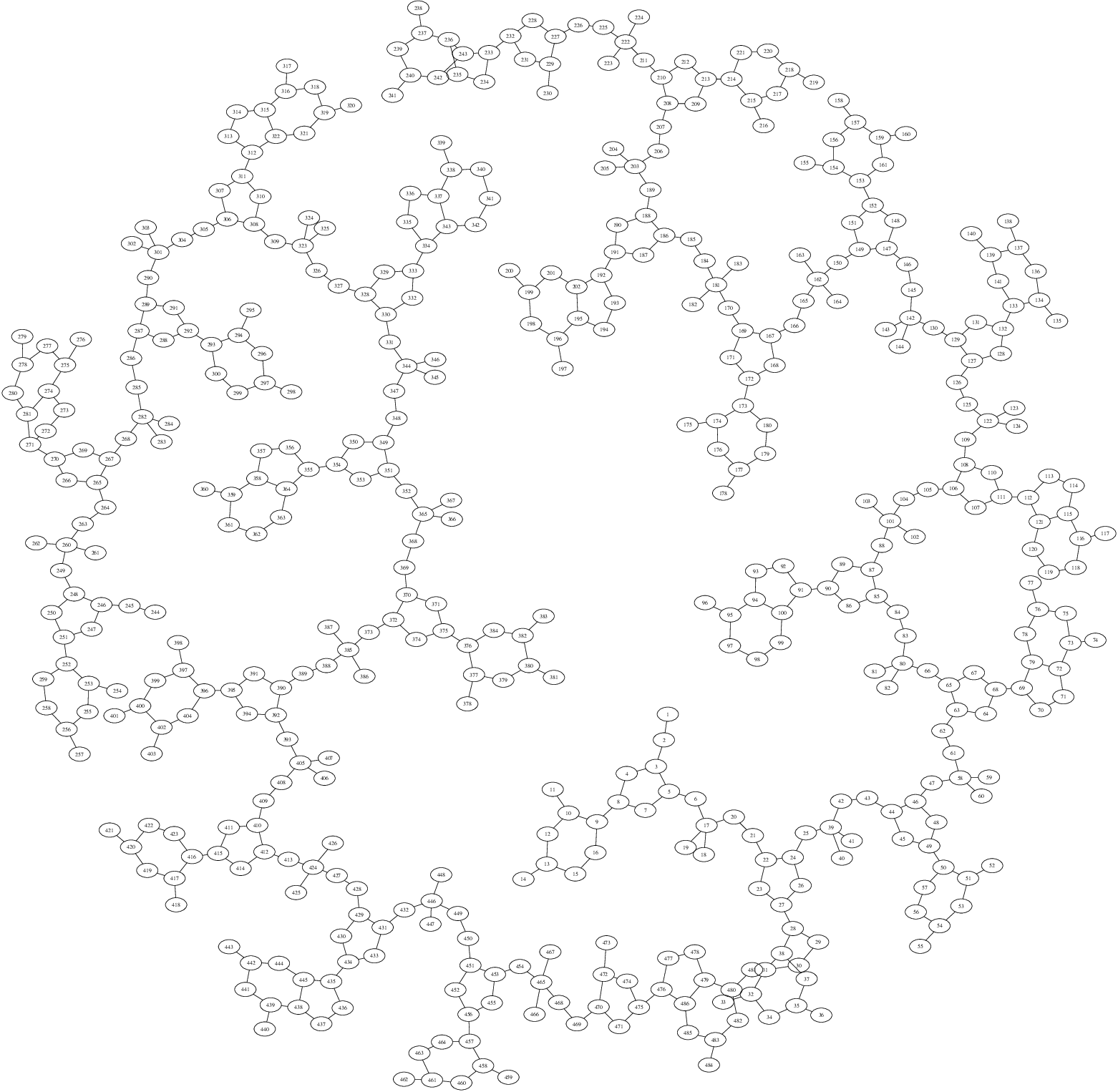}
}
\caption{
Layout of original graph corresponding to a small DNA molecule.
}\label{DNA}
\end{figure}

\begin{figure}[bt]
\centerline{\includegraphics[height=7cm]{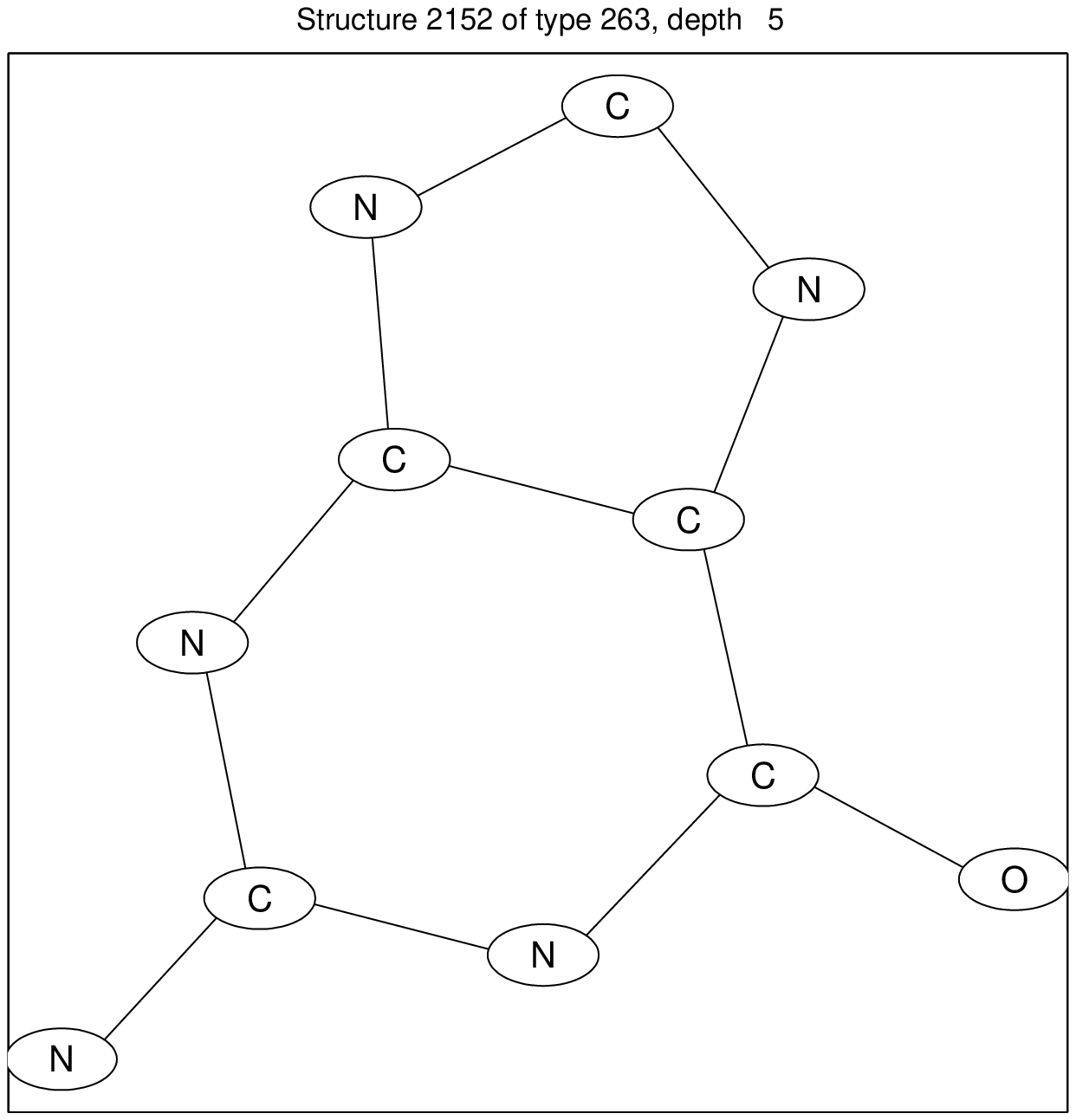}{\includegraphics[height=7cm]{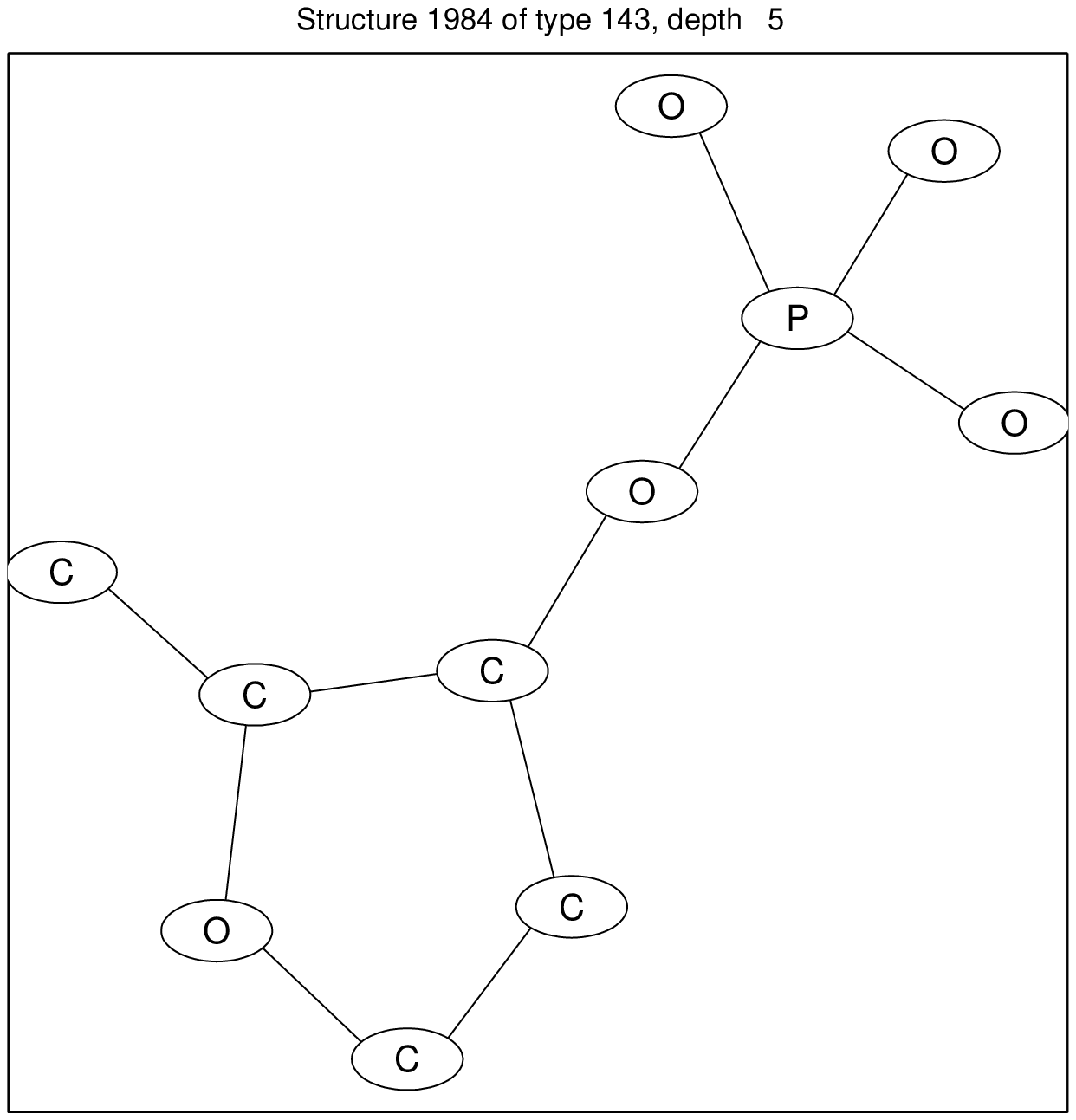}}}
\caption{{\bf left:} The compound object induced by the {\em Graphitour} algorithm, which corresponds to the base of Guanine. All four DNA bases were identified 
as non-terminal nodes in the resulting graph grammar.
{\bf right:} The compound object induced by the {\em Graphitour} algorithm, which
corresponds to the backbone of the molecule: phosphate and sugar}\label{figX}
\end{figure}

In this section we take a closer look at the results of running {\em Graphitour}
on raw data obtained from a PDB file of a DNA molecule by stripping
off all the information except for the list of atoms and the list of of covalent bonds.  To simplify the representation of results, hydrogen atoms and corresponding edges were removed from the graph. Additionally, to support the symmetry, we artificially linked the DNA into a loop by extra bonds from 3' to 5' ends. 
The resulting list of $776$ nodes and $2328$ edges in the incidence matrix
is presented as input to {\em Graphitour} and is depicted in Figure~\ref{DNA}.
Figure~\ref{figX}.left and ~\ref{figX}.right show two sample non-terminals 
from the actual output from the {\em Graphitour} software, illustrating the
discovery of basic DNA elements. 

Figure~\ref{figX}.left shows the compound object induced by the {\em Graphitour}
algorithm, which corresponds to the base of Guanine. All four DNA bases 
were identified as non-terminal nodes in the resulting graph grammar.
This particular structure is identified as a non-terminal by the number
$2152$ in a list of non-terminals and described as type $263$ in the
lexicon of compounds. It is at the depth $5$ in the hierarchy tree of
the whole graph, being joined with the corresponding backbone phosphate
\& sugar one level higher in the upstream;
subsequently joined with the complete nucleotide adjacent to it, and so forth.

Figure~\ref{figX}.right shows the compound object induced by the {\em Graphitour}
algorithm, which corresponds to the backbone of the molecule: joined phosphate
and sugar. This particular structure is identified as a non-terminal by 
the number $1984$ in a list of non-terminals and described as type
$143$ in the lexicon of compounds. It is at the depth $5$ in the
hierarchy tree of the whole graph, same depth as compound in
Figure~\ref{figX}.left which makes sense. Note that although the number of 
lexicon entries is over a hundred, the vast majority of numbers were only used for
intermediate entries, which got merged into composite elements. There is no reusing of the available lexicon slots. Such re-use could give some extra compression efficiency (compare to {\em Sequitour}~\cite{Sequitour}). The resulting lexicon is small and all the entries in the final lexicon are biologically relevant compounds. 

{\em Graphitour} actually discovers the hierarchical structure corresponding
to the kind of zoom-in hierarchical description we gave in the introduction. The algorithm identifies all of these structural elements without exception and does not select any elements which would not make biological sense. The 
significance of such result might not be immediately clear if one does
not take into account that the algorithm does not have any background
knowledge, any initial bias, nothing at all except for the list of
nodes and edges.

\begin{figure}[bth]
\centerline{\includegraphics[width=6cm]{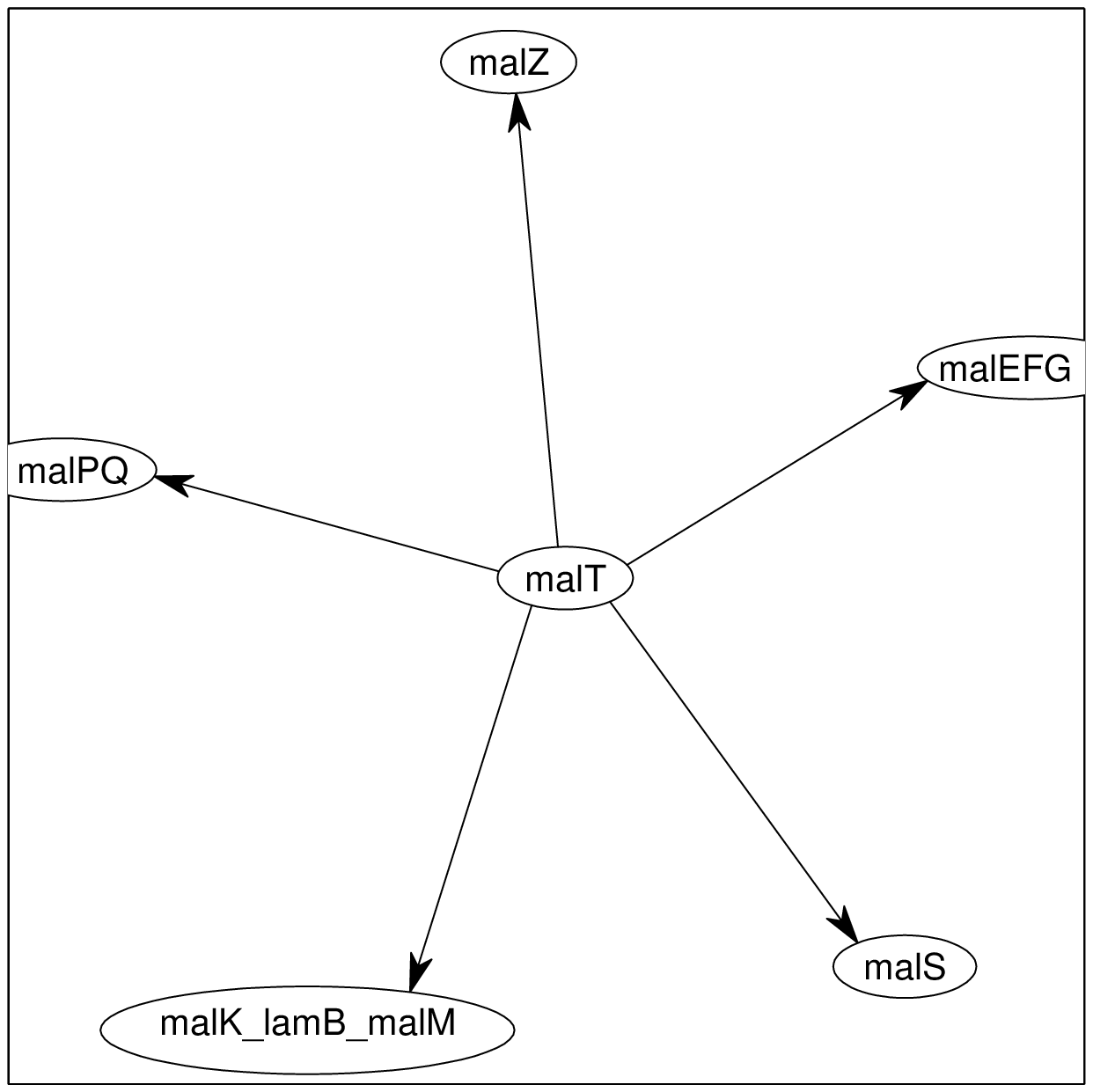}\includegraphics[width=6cm]{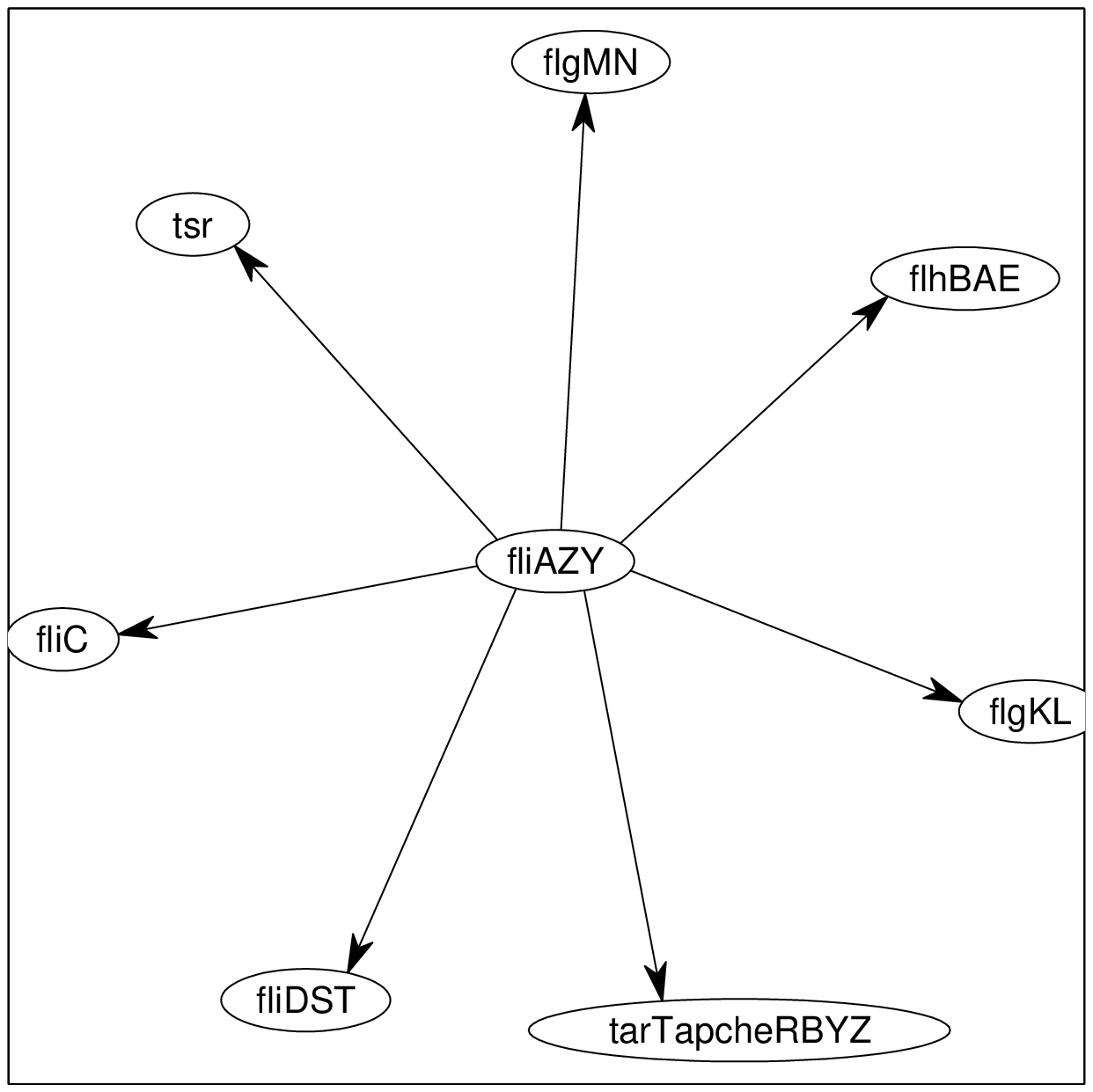}}
\centerline{\includegraphics[width=6cm]{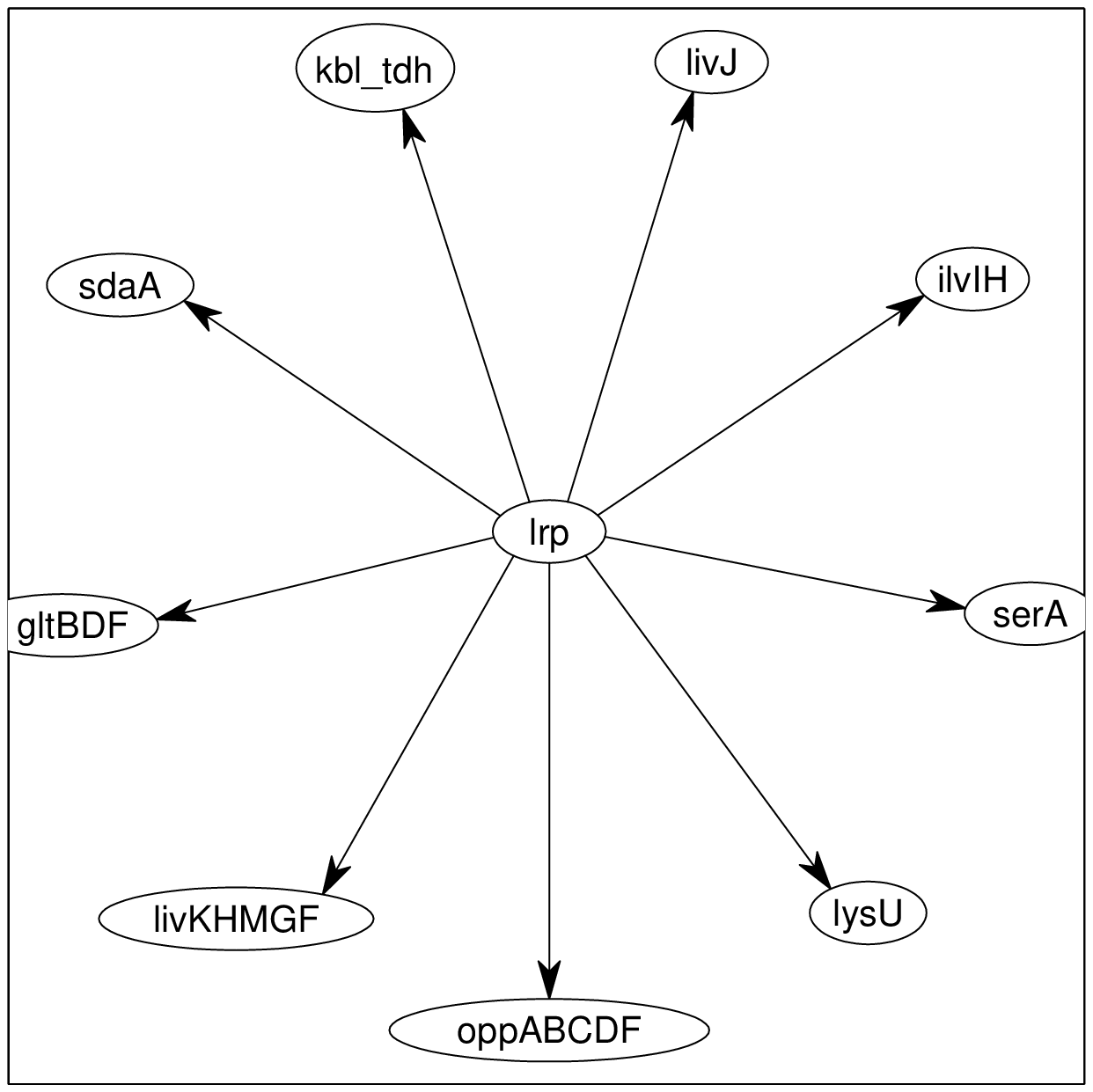}\includegraphics[width=6cm]{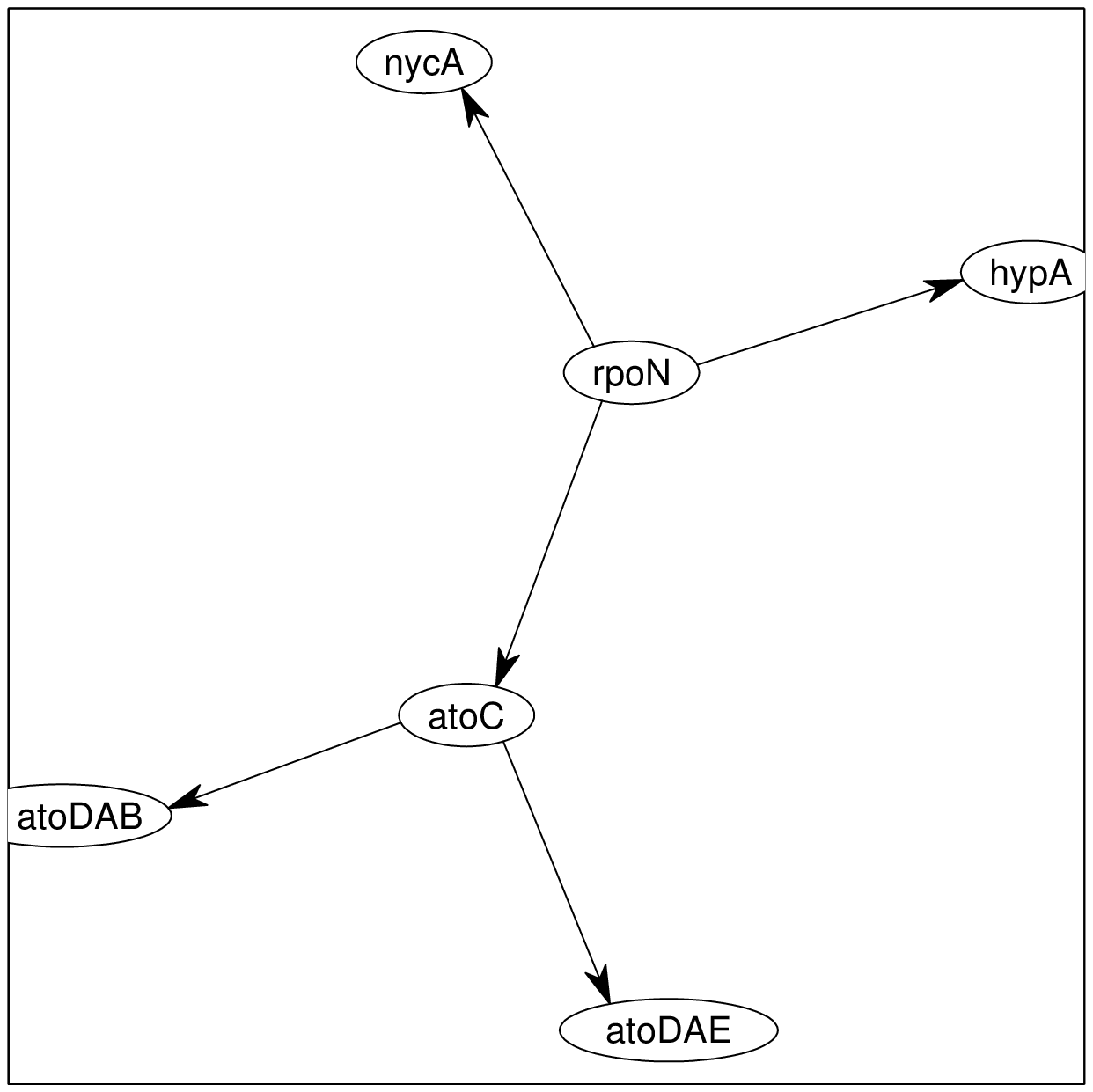}}
\caption{Compositional elements of E.coli regulatory network.}\label{alon} 
\end{figure}

As an example of a rather different application domain, we tried {\it Graphitour} on {\it escherichia coli} transcriptional regulation network, since it is one of the well known attempts to discover "network motifs" or building blocks in a graph as over-represented sub-graphs by Alon et al.~\cite{Alon02}. In this network there are  $423$ nodes corresponding to the genes or groups of genes jointly transcribed (operons). Edges correspond to the $577$ known interactions: 
each edge is directed from an operon that encodes a transcription factor to an operon that it directly regulates. For more information on the biological significance of the motif discovery in regulatory networks we refer you to the cited paper~\cite{Alon02} and supplementary materials at the Nature Internet site. 
Alon et al. find that much of the network could be composed of repeated appearances of three highly significant motifs, that is a sub-graph, of a fixed topology, in which nodes are instantiated with different node labels. One example is what they call a feedforward loop $X\rightarrow Y \rightarrow Z$ combined with $X \rightarrow Z$, where $X,Y,Z$ are instantiated by particular genes. 
In order to compare to the results of Alon et al. we label nodes simply according to cardinality, omitting gene identities, and take into account directionality of the edges. No hierarchical structure or feedforward loops were distilled from the data. Rather, {\it Graphitour} curiously disassembled the network back into the sets of star-like components as illustrated by Figure~\ref{alon}, which ought to have been initially put together to create such network. It leaves open the question of what is a useful definition of a network component in biological networks, if not the one requiring the most parsimonious description of the network. 

\section{Implementation}

The implementation of this algorithm was done in Matlab version 6.5 and relies on two other important components also developed by the author. The first is a maximum cardinality non-bipartite matching algorithm by Gabow~\cite{Gabow73} implemented by the author in Matlab. The other component is a graph layout and plotting routine. Representing the original input graph as well as the resulting graph grammar is a challenging task and is well outside the scope of this paper. For the purposes of this study we took advantage of existing graph layout package from AT\&T called GraphViz~\cite{GraphViz}. The author has also developed a library for general Matlab-GraphViz interaction, i.e. converting the graph structure into GraphViz format and converting the layout back into Matlab format for visualization from within Matlab.  
Supplementary materials available upon request contain several screen-shots of executing the algorithm on various sample graphs, including regular rectangular and hexagonal grids, synthetic floor plans and trees with repetitive nested branch structure. These are meant to both illustrate the details of our approach and provide a better intuition of iterative analysis.

\section{Discussion}

We presented a novel algorithm for the induction of hierarchical structure in labeled graphs and demonstrated successful induction of a multi-level structure of a DNA molecule from raw data---a PDB file stripped down to the list of atoms and covalent bonds. 

The significance of this result is that the hierarchical structure inherent in a multi-level nested network can indeed be inferred from first principles with no domain knowledge or context required. It is particularly surprising that the result is obtained in the absence of any kind of functional information on the structural units. Thus, {\em Graphitour} can automatically uncover the structure of unknown complex molecules and ``suggest'' functional units to human investigators. 
While, there is no universal way to decide which components of the resulting structure are "important" across all domains and data, frequencies of occurrence is one natural indicator. 

One exciting area of application which would naturally benefit from {\em Graphitour} is automated drug discovery. Re-encoding large lists of small molecules from PBD-like format into grammar-like representation would allow for a very efficient candidate filtering at the top levels of hierarchy with the added benefit of creating structural descriptors to be matched to functional features with machine learning approaches. 

Analysis of proteins benefits from the same multi-level repetition of structure as DNA, and {\em Graphitour} gets similar results for individual peptides. Analyzing a large set of protein structures together is our work in progress. The hope is to reconstruct some of the domain nomenclature and possibly re-define the hierarchy of the domain assembly. 

The current version of the {\em Graphitour} corresponds to a so-called non-lossy or lossless compression. This guarantees that the algorithm will induce a grammar corresponding precisely to the input data, without over- or under-generating. The drawback of this approach is that the algorithm is sensitive to noise and has rather poor abstraction properties. Future work would require developing lossy compression variants, based on the approximation of description length minimization. 

Finally, it would be quite interesting to analyze the outcome of the algorithm from a cognitive plausibility point of view, that is to compare structures found by human investigator to these descovered by the algorithm on various application examples. 

\section*{Acknowledgement}
A substantial part of this work was done while the author was visiting MIT CSAIL with Prof. Leslie Kaelbling, supported by the 
DARPA through the Department of the Interior, NBC, Acquisition Services Division, under Contract No. NBCHD030010. 


 \bibliographystyle{plain}
{\footnotesize

\end{document}